# Mechanical responses of borophene sheets: A first-principles study

Bohayra Mortazavi[*,1], Obaidur Rahaman[r,1], Arezoo Dianat[2], Timon Rabczuk[#,1]

[1]Institute of Structural Mechanics, Bauhaus-Universität Weimar, Marienstr. 15, D-99423 Weimar, Germany.

[2]Institute for Materials Science and Max Bergman Center of Biomaterials, TU Dresden, 01062 Dresden, Germany

## Abstract

Recent experimental advances for the fabrication of various borophene sheets introduced new structures with a wide prospect of applications. Borophene is the boron atoms analogue of graphene. Borophene exhibits various structural polymorphs all of which are metallic. In this work, we employed first-principles density functional theory calculations to investigate the mechanical properties of five different single-layer borophene sheets. In particular, we analyzed the effect of loading direction and point vacancy on the mechanical response of borophene. Moreover, we compared the thermal stabilities of the considered borophene systems. Based on the results of our modelling, borophene films depending on the atomic configurations and the loading direction can yield remarkable elastic modulus in the range of 163-382 GPa.nm and high ultimate tensile strength from 13.5 GPa.nm to around 22.8 GPa.nm at the corresponding strain from 0.1 to 0.21. Our study reveals the remarkable mechanical characteristics of borophene films.

*Corresponding author (Bohayra Mortazavi): bohayra.mortazavi@gmail.com
Tel: +49 157 8037 8770,
Fax: +49 364 358 4511
[r] amieor@gmail.com
[#]timon.rabczuk@uni-weimar.de

## 1. Introduction

Over the last couple of decades, there have been tremendous efforts for the fabrication of two-dimensional (2D) materials. 2D materials are currently considered among the most interesting research topics because of their remarkable and wide potential applications. The interest toward this new class of materials was originated by the successful production of graphene[1–3], the honeycomb lattice of carbon atoms in planar form. Graphene is a semi-metallic material which exhibits exceptional mechanical[4] and heat conduction[5] properties. After the synthesis of graphene, other 2D materials were successfully fabricated such as hexagonal boron-nitride[6,7], graphitic carbon nitride[8], silicene[9,10], germanene[11], stanene[12] and transition metal dichalcogenides[13–15] like molybdenum disulfide ($MoS_2$). Nevertheless, the interest toward the synthesis and application of 2D materials sounds to be stopples. In line with the continuous advances in the fabrication of new 2D materials, exciting developments have just taken place with respect to the synthesis of borophene[16,17]. To the best of our knowledge, three different 2D boron films have been so far experimentally fabricated all by epitaxial growth of boron atoms on silver substrate[16,17]. In accordance with theoretical predictions[18,19], these sheets present metallic properties. For the real applications of these new films, hence a comprehensive understanding of their properties plays a critical role. In this regard, theoretical studies can be considered as promising approaches to assess the properties of these materials that are difficult, expensive and time consuming to be experimentally evaluated[20–28]. One of the key factors for the application of a material is its mechanical properties that correspond to the stability of the material under the applied mechanical strains. In this work we therefore studied the mechanical properties of five different borophene sheets using first-principles density functional theory (DFT) calculations. We elaborately studied the effect of loading direction and defects formation on the mechanical properties of different borophene sheets.

## 2. Atomistic modelling

In this study, we investigated the mechanical properties of five different boron sheets (S1-S5) which are illustrated in Fig. 1. To probe the effect of loading direction, we analyzed the mechanical properties along armchair and zigzag directions as depicted in Fig. 1. All constructed samples were periodic in the planar directions therefore the obtained results correspond to infinite sheets and not the borophene nanoribbons. The sheet S1 is the densest structure with out of plane buckling which was synthesised experimentally by Mannix *et al.*[16]. The sheets S2 and S3 are planar borophene sheets that have been most recently experimentally fabricated by Feng *et al.*[17]. In addition we considered two other sheets that have been theoretically predicted[29]. In this study, we used relatively large atomic models with 75 atoms to 128 atoms. DFT calculations were performed as implemented in the Vienna ab initio simulation package (VASP)[30,31] using the Perdew-Burke-Ernzerhof (PBE) generalized gradient approximation exchange-correlation functional[32]. The projector augmented wave method [33] was employed with an energy cutoff of 500 eV. Conjugate gradient method was used for the geometry optimizations. Periodic boundary conditions were applied in all direction and a vacuum layer of 20 Å was considered to avoid image-image interaction along the sheet thickness. For the evaluation of mechanical properties, Brillouin zone was sampled using a 5×5×1 k-point mesh size and for the calculation of electronic density of states we performed a single point calculation in which the Brillouin zone was sampled using a 11×11×1 k-point mesh size using Monkhorst-Pack mesh[34]. We applied uniaxial tension condition to evaluate the mechanical properties of various borophene sheets. To this aim, we increased the periodic simulation box size along the loading direction in multiple steps with a small engineering strain steps of 0.003. In this case, when the system is stretched in one direction, the stress on the perpendicular direction may not be zero. Depending on the Poisson's ratio of the material, stretching the structure along the one direction cause stretching (Poisson's ratio>0) or contracting (Poisson's ratio<0) stresses on the perpendicular directions. Since we deal with planar materials, the atoms are in contact with vacuum along the thickness. For example, for a 2D material when the structure is stretched along the armchair direction, there might be stresses along the zigzag direction. For this condition, to ensure accurate uniaxial stress condition, the simulation box size along the perpendicular direction of the loading was changed in a way that the stress in this direction remained negligible in comparison with that along the loading direction. The atomic positions were accordingly rescaled according to the changes in the simulation box size. Finally, conjugate gradient method was used for the geometry optimizations. Worthy to note that at every step of loading, small random displacements along the planar direction were added to the boron atoms positions to avoid error in the VASP calculation due to the symmetrical atomic positions. For the ab initio molecular dynamics (AIMD) simulations, Langevin thermostat was used for maintaining the temperature using a time step of 1fs. In this case we used 2×2×1 k-point mesh.

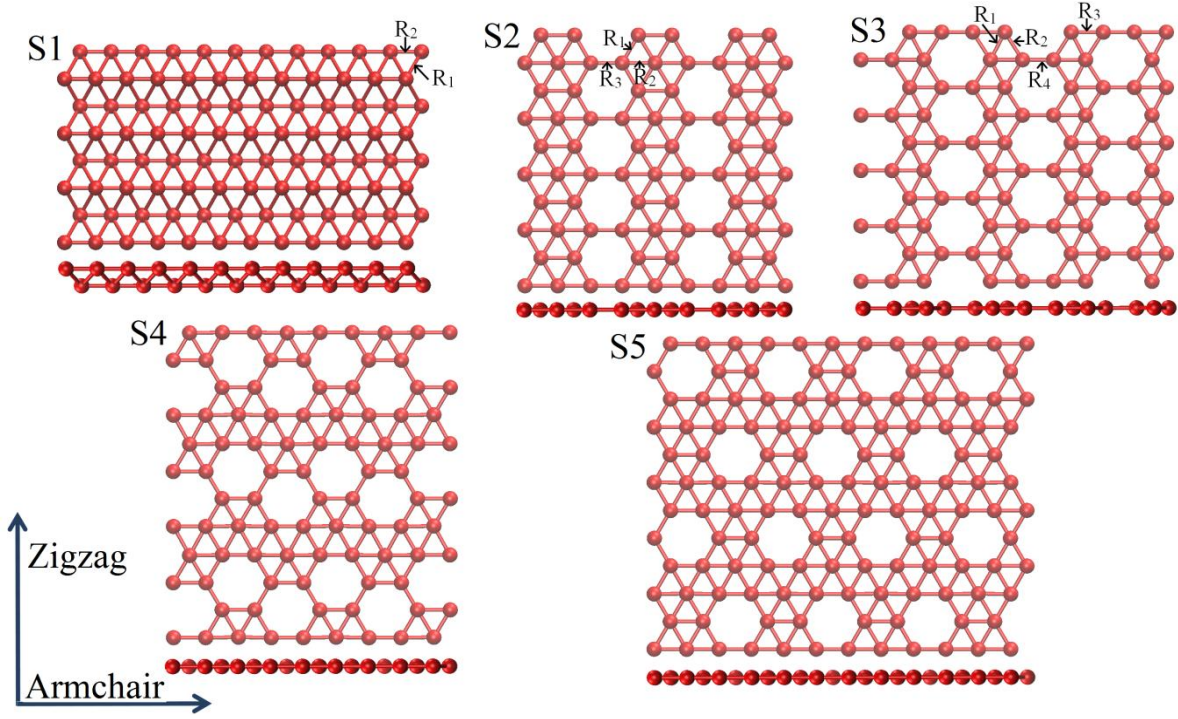

Fig. 1- Top and side views of five different borophene sheets (S1-S5) constructed in this study. The mechanical properties are studied along armchair and zigzag directions. VMD [35] software was used to illustrate the structures

## 3. Results and discussions

In Fig. 2, the deformation process of borophene sheets stretched along the armchair direction are depicted. We illustrated the structures at three different strain levels ($\varepsilon$): energy minimized structure ($\varepsilon=0.0$), under half of the strain at ultimate tensile strength ($\varepsilon = 0.5\varepsilon_{uts}$) and finally at ultimate tensile strength point. For the S1 borophene, due to the presence of the most regular atomic structure we observe an uniform extension of the sample along the loading direction. In this case, the buckling length decreased gradually with increasing strain level. For the S2 and S3 structures, the stretching did not occur uniformly. For these sheets the bonds connecting hexagonal (in S2 borophene) or zigzag (in S3 borophene) lattices are along the loading direction and they stretched more considerably in comparison with other bonds. In S4 borophene, we also found higher deformation for the bonds connecting fully occupied hexagon lattices. Nevertheless since these bonds are not along the loading direction their deformation is closer to the other bonds in the structure. Regarding the S5 borophene, we observed higher stretching of the bonds around the hexagonal holes in the structure. Apart from the S1 films, increasing the strain levels along the loading direction decreased the periodic sheet size along the transverse direction. For small strain levels within the elastic regime, strain along the traverse direction ($\varepsilon_t$) with respect to the loading strain ($\varepsilon_l$) is acceptably constant. In this case one can evaluate the Poisson's ratio by calculating the: $-\varepsilon_t/\varepsilon_l$. For S2 to S5 borophene films this ratio was positive meaning that the structures shrank in one direction when they were stretched along the other direction. Nevertheless, for the S1 borophene we observed very slight increase in the periodic simulation box size in perpendicular direction to the loading direction which indicated a small negative Poisson's ratio for this structure.

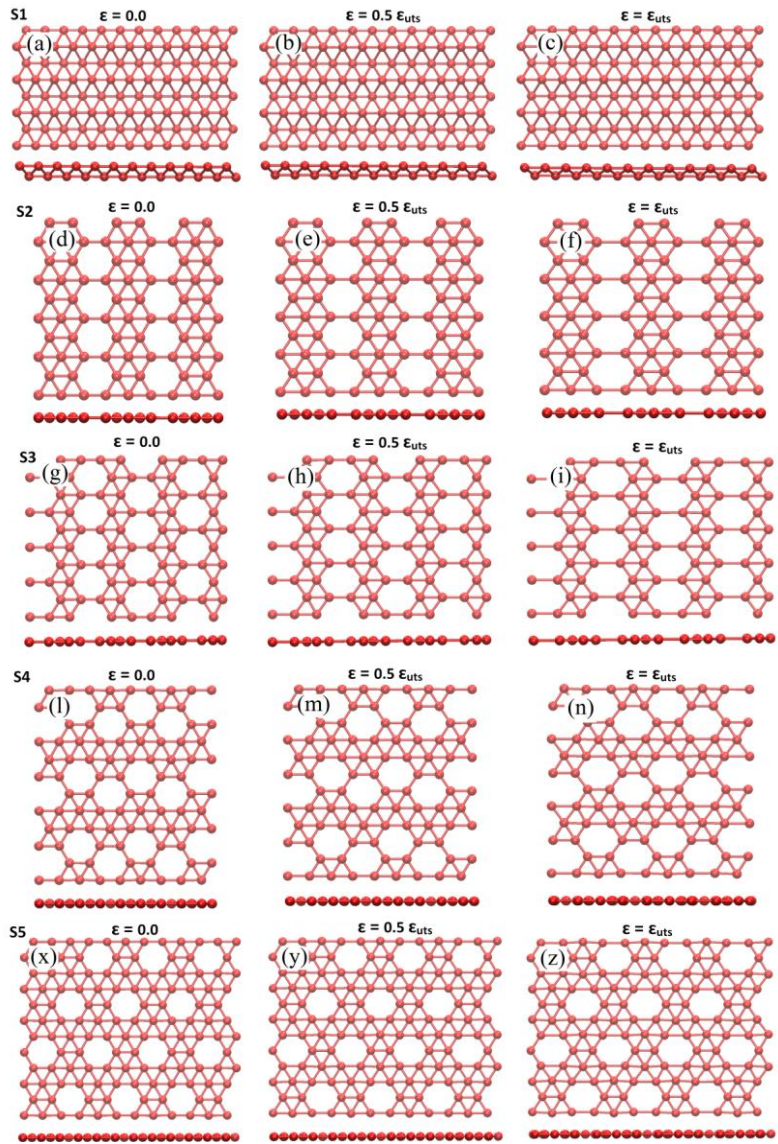

Fig. 2- Top and side view of deformation processes of single-layer borophene sheets for different strain levels ($\varepsilon$) with respect to the strain at ultimate tensile strength ($\varepsilon_{uts}$).

Calculated uniaxial stress-strain responses of defect-free and single-layer borophene films along armchair and zigzag loading directions are illustrated in Fig. 3. In all the cases, the stress-strain curves include an initial linear relation which is followed by a nonlinear trend up to the ultimate tensile strength at which the material yields its maximum load bearing ability. After the ultimate tensile strength point the stress decreases by increasing the strain level. The strain at which the ultimate tensile strength occurs is also an important parameter which identifies how much the material can be stretched before reducing its load bearing ability. The stress-strain responses of borophene sheets correlate with their atomic configurations and the way they evolve and rearrange during the loading condition. For example, for S2 and S3 borophene films along the armchair direction, the fully occupied hexagonal or zigzag lattices are connected by single B-B bonds and therefore the stretching limit of these bonds plays the critical role in determining the ultimate tensile strength point of these structures. As it is shown in Fig. 3, along the armchair direction both S2 and S3 films yield very close ultimate tensile strength points. On the other hand, when these sheets are stretched along the zigzag direction, the fully occupied hexagonal or zigzag lattices are along the loading direction and their stretching characteristics define the ultimate tensile strength point of the structure. Based on our simulations fully occupied hexagonal lattices in S2 borophene can extend more than zigzag lattices in S3 borophene.

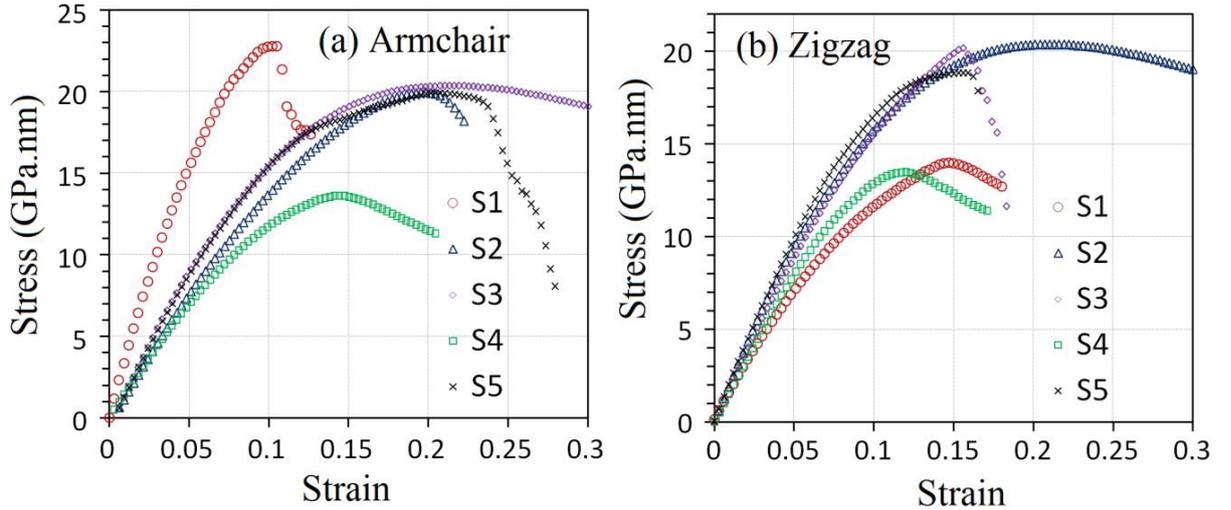

Fig. 3- Calculated uniaxial tensile stress-strain response of defect-free and single-layer borophene films along (a) armchair and (b) zigzag loading directions.

Table 1, Mechanical properties of borophene sheets, Y, P, STS and UTS depict elastic modulus, Poisson's ratio, strain at ultimate tensile strength point and ultimate tensile strength, respectively. Stress units are in GPa.nm.

| Structure | $Y_{armchair}$ | $Y_{zigzag}$ | P | $STS_{armchair}$ | $STS_{zigzag}$ | $UTS_{armchair}$ | $UTS_{zigzag}$ |
|---|---|---|---|---|---|---|---|
| S1 | 382 | 163 | -0.01 | 0.105 | 0.145 | 22.8 | 14 |
| S2 | 190 | 210 | 0.18 | 0.2 | 0.21 | 19.97 | 20.38 |
| S3 | 208 | 205 | 0.11 | 0.21 | 0.155 | 19.91 | 20.18 |
| S4 | 186 | 167 | 0.26 | 0.14 | 0.12 | 15.65 | 13.48 |
| S5 | 214 | 217 | 0.2 | 0.21 | 0.16 | 14.84 | 18.83 |

The mechanical properties of borophene sheets predicted by our DFT calculations are summarized in Table 1. We note that the elastic modulus in the present study was evaluated by fitting a straight line to the stress-strain curves for the strain level up to 0.006. The elastic modulus of the considered structures were similar when stretched along armchair direction or the zigzag direction, with an exception for the S1 borophene. According to our calculations for the S1 structure, the elastic modulus was found to be 382 GPa.nm along the armchair direction and 163 GPa.nm along the zigzag direction. Our predictions for this borophene membrane are slightly below the previously reported elastic modulus of 398 GPa.nm[16] and 389 GPa.nm[36] along the armchair and 170 GPa.nm[16] and 166 GPa.nm[36] along the zigzag directions. This elastic modulus anisotropy for S1 graphene can be explained because of its structural features. Interestingly, the elastic modulus of S1 structure along the zigzag direction is the lowest among the studied samples in the present study. A comparison of the strain at ultimate tensile strength values suggests the lowest value for S1 borophene stretched along the armchair direction (around 0.1). On the other hand, the S2 structure when stretched along the zigzag direction presents the highest strain at ultimate tensile strength (about 21%). A comparison of the ultimate tensile strength values suggests the highest of about 23 GPa.nm GPa for S1 structure when stretched along the armchair direction. S4 borophene when stretched along the zigzag direction yields the lowest tensile strength of 13.48 GPa.nm. The Poisson's ratio of S1 structure was found to be negative and close to zero. The Poisson's ratios of the other four structures ranged from about 0.1 to 0.25. We found that the Poisson's ratio of borophene films are convincingly independent of the loading direction. We note that according to recent theoretical predictions[36,37], S1 borophene film present phonon instability likely to $MoS_2$[36]. It was concluded that the S1 borophene lattice may exhibit instability against long-wavelength transversal waves[36]. The investigation of phonon instability of considered borophene films under different loading conditions is therefore an interesting topic for the future studies.

Fig. 4 shows the samples of B-B bond lengths evolution during the uniaxial stretching of S1, S2 and S3 borophene sheets along the armchair direction. As expected, the bonds that were along the direction of stretching were gradually elongated with increasing strain levels. For all of the studied sheets, we found that the bonds that were not along the direction of stretching either decreased in length (e.g. Fig. 4a, $R_2$ bond) or remained almost similar to their initial values (e.g. Fig. 4b, $R_1$ bond) with increasing the strain levels. In the cases of S2, S3, S4 and S5 borophene membranes non-identical stretching were observed for different bond types that were along the direction of stretching. For example, for the S2 sheet, $R_3$ bond (as depicted in Fig. 1) was stretched at a higher rate than $R_2$ bond, in a similar way, for the S3 sheet, $R_4$ bond was stretched at a higher rate than the $R_3$ bond. These can probably be explained by the higher propensity of the two center-two electron (c2-2e) bonds ($R_3$ in the case of S2 and $R_4$ in the case of S3) to be stretched easier as compared to the three center-two electron (3c-2e) bonds ($R_2$ in the case of S2 and $R_3$ in the case of S3).

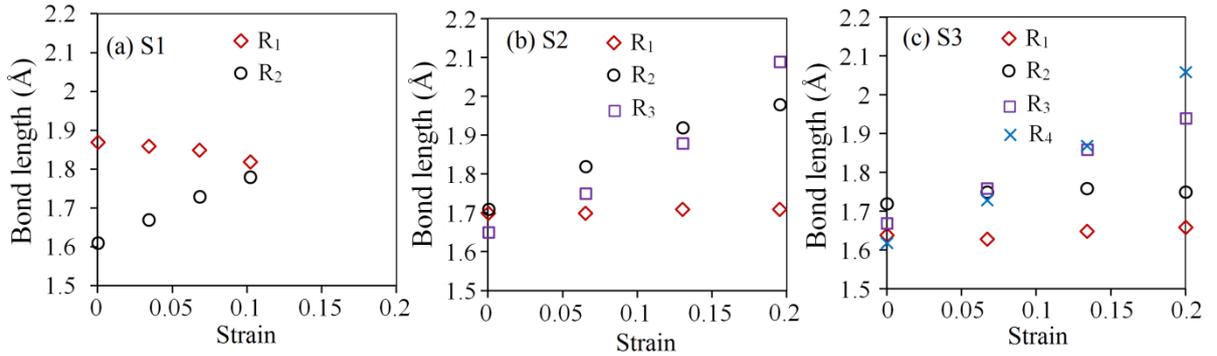

Fig. 4- Samples of variation of bond lengths for various strain levels for structures (a) S1, (b) S2 and (c) S3, uniaxially loaded along the armchair direction. Different bonds $R_1$, $R_2$, $R_3$ and R4 for studied systems are depicted in Fig. 1.

All five considered borophene sheets at different strain levels up to the tensile strength point were selected for electronic density of states (DOS) calculation. Fig. 5 illustrate samples of acquired DOS curves for borophene films elongated along the armchair direction. As it can be observed, in the calculated total DOS for the relaxed and uniaxially loaded systems, at the zero state energy (Fermi level) the DOS is not zero which consequently demonstrate metallic behaviour. For the S1 borophene, we found that upon the stretching, the total DOS for valence/conduction band around the Fermi level are increased. On the other hand, for the rest of considered borophene films we observe that during the elongation the total DOS around the Fermi level slightly decrease. Nevertheless, according to our finding, one cannot open a band gap by stretching the borophene films.

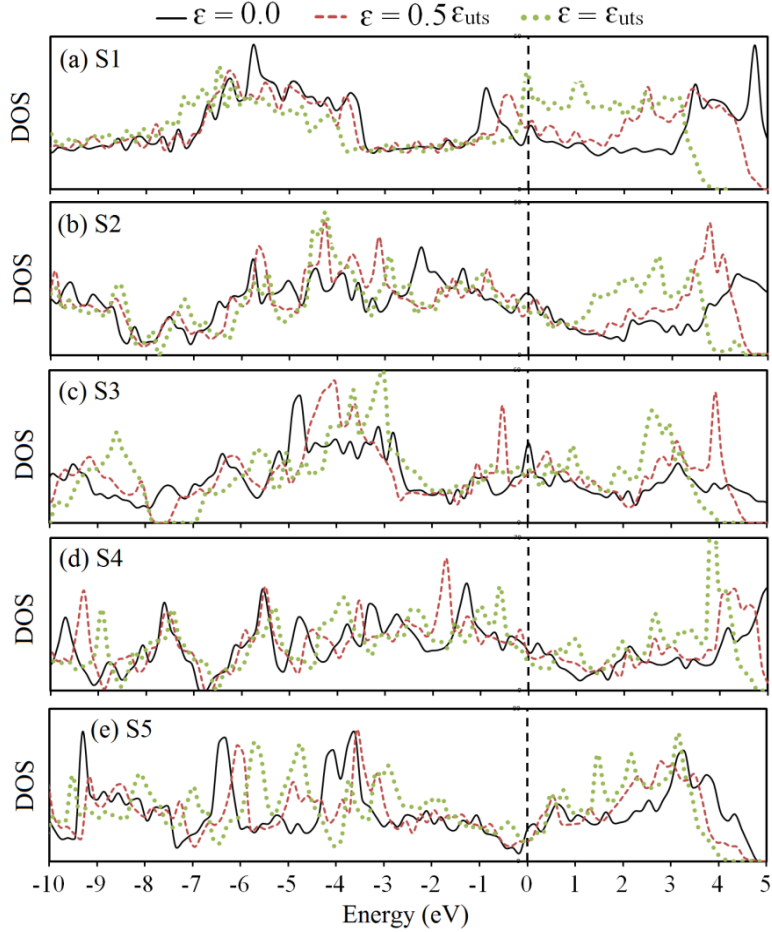

Fig. 5- Total electronic density of states (DOS) for borophene structures elongated along the armchair directions at different strains ($\varepsilon$) with respect to the strain at ultimate tensile strength ($\varepsilon_{uts}$).

Like all other known materials, experimentally fabricated borophene sheets are expected not to be perfect and different types of defects may exist in borophene lattices. Defects in materials influence both physical and chemical properties and in some cases it may substantially affect the electronic properties[38–40]. Defects in 2D materials such as graphene are formed mainly during the fabrication possess. Despite the high thermal stability of graphene, with a melting point of 4510 K [41], crystal growth during chemical vapour deposition (CVD) technique form various types of defects in graphene [40,42]. Since the thermal stability of borophene sheets are considerably lower than graphene, the existence of defects in experimentally fabricated borophene sheets are therefore expected to be common. The presence of defects can naturally affect the mechanical properties of 2D materials. In order to understand the role of defects on the mechanical properties of borophene sheets, we introduced point defects by gradually removing 1, 2, 3 or 4 atoms from the structures. We note that more than 10 different borophene films have been predicted theoretically[29] or fabricated experimentally[16,17]. The difference of these boron films are due to the regular pattern of the removed boron atoms in the unit-cell. This way, for the modelling of defective borophene films we accordingly removed the boron atoms randomly such that the obtained structure is not symmetrical and similar to the other borophene films. After removing few boron atoms for every structure, we performed energy minimization to obtain the relaxed structure.

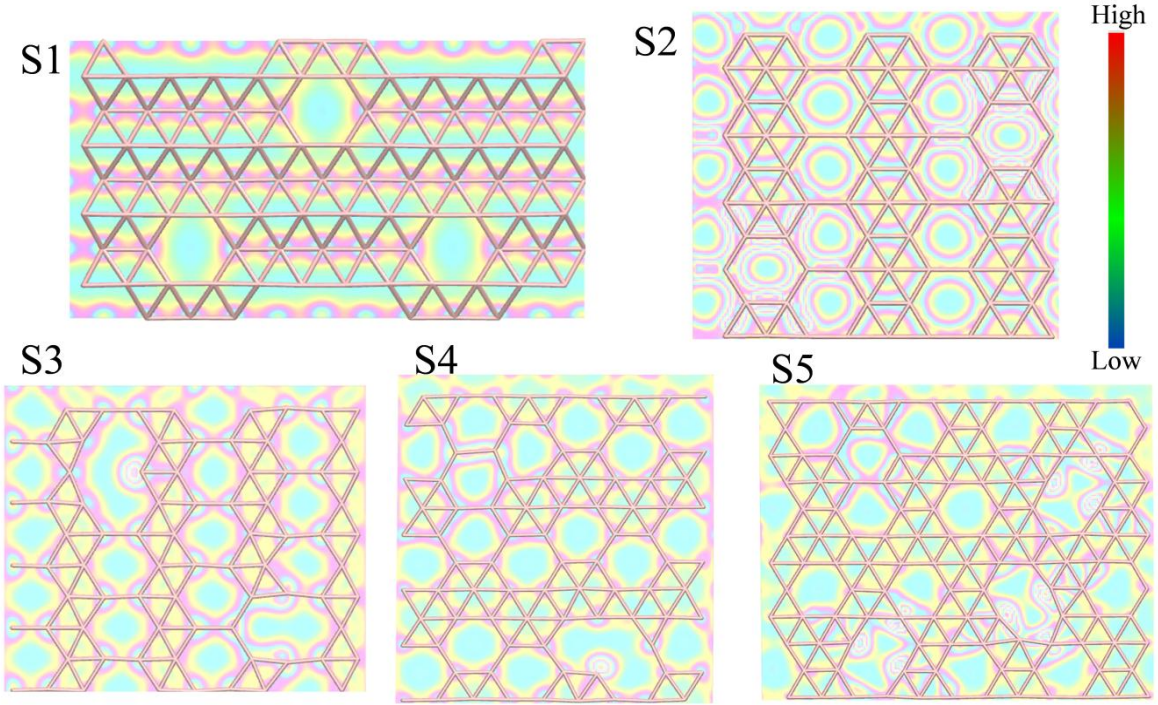

Fig. 6- Top view of various minimized defective borophene sheets. The contour illustrate partial charges density around the Fermi energy. For the S1 sheet, the charge density is plotted for the atoms that are placed on the botton atomic plane.

In Fig. 6, the top views of the minimized highly defective borophene sheets are illustrated. We observed remarkable out of plane deflection of S1 sheets whereas the other structures are kept planar. In order to analyze the correlation between electronic and mechanical properties of defective structures, we plotted the partial charges density around the Fermi energy for the borophene films with highest defect concentrations in Fig. 6. The charge localization effect determines the elastic properties of the nanostructures. Higher charge localization leads to stronger bonding energy and consequently higher mechanical stability[43]. Highest localized charges are observed between B-B bonds for S1 structure resulting in the highest elastic modulus for this structures. The lowest charge density between B-B bonds are indicated for S2 borophene. Because of the similar charge distributions for S3, S4, and S5 structures the correlation between charge density and the elastic modulus is not trivial. However, all these structures present close elastic modulus. In addition, we conducted electronic density of states calculation for constructed defective borophene sheets. Our calculated total DOS confirm that in all cases, the structures present metallic behaviour as indicated by the lack of any band gaps in the DOS.

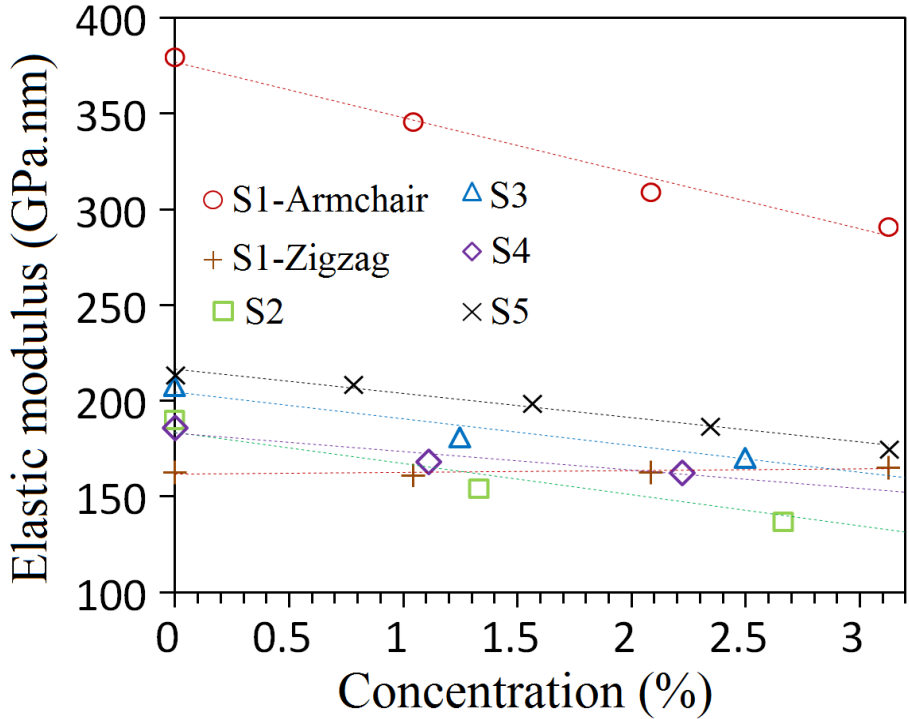

Fig. 7- Elastic modulus of borophene sheets as a function point defect concentration.

After obtaining the minimized structures, the borophene films were subjected to loading strains to evaluate the elastic modulus. Since the elastic modulus of S1 borophene is highly anisotropic, in this case we calculated the elastic modulus along both armchair and zigzag directions. Elastic modulus of borophene sheets as a function of point defects concentration are illustrated in Fig. 7. Based on our simulations, the elastic modulus of considered borophene films along the armchair direction decreased almost linearly with increasing defects concentration. It is worthy to note that based on classical molecular dynamics simulations it was predicted that the elastic modulus of graphene decreases also linearly with increasing the defects concentrations [44,45] and such a relation was found to be consistent also for amorphized graphene [45]. The decreasing trends in the elastic modulus and tensile strength of a covalently bonded material by increasing the defects concentration is expected since the loss of an atom not only removes bonds that are involved in the load transfer but also causes stress concentrations which consequently reduce the mechanical strength. Interestingly, our first principles calculation reveals that the elastic modulus of S1 structure along the zigzag direction rather slightly increases with increasing the defects concentration. To understand the mechanism behind such an unexpected trend, we should remember that the sheets S2 to S5 are nothing but the structure S1 in which some atoms are removed with special patterns leading to the formation of planar films. Based on our results discussed earlier, the elastic modulus of these structures are all higher than that for original S1 sheet along the zigzag direction. So one can conclude that by increasing the defects concentrations in S1 borophene its structure will approach other planar and pristine borophene sheets. Since the elastic modulus along zigzag direction for these borophene sheets are higher than that for the S1 borophene along zigzag direction, by increasing the defect concentration the S1 borophene present slightly higher elastic modulus along the zigzag direction. Worthy to note that such a trend is predicted to be valid only for small defects concentrations and after a point, further increasing of the defect concentrations will result in a decline in the elastic modulus. For the defective films, we also calculated the Poisson's ratio and we found that by increasing the defect concentration the Poisson's ratio does not change considerably. Only for the S1 borophene, we found that by increasing the defect concentration the Poisson's ratio become slightly positive. This finding was expectable because of the fact that S1structure by increasing the defect concentration approaches the other borophene films which present positive Poisson's ratio.

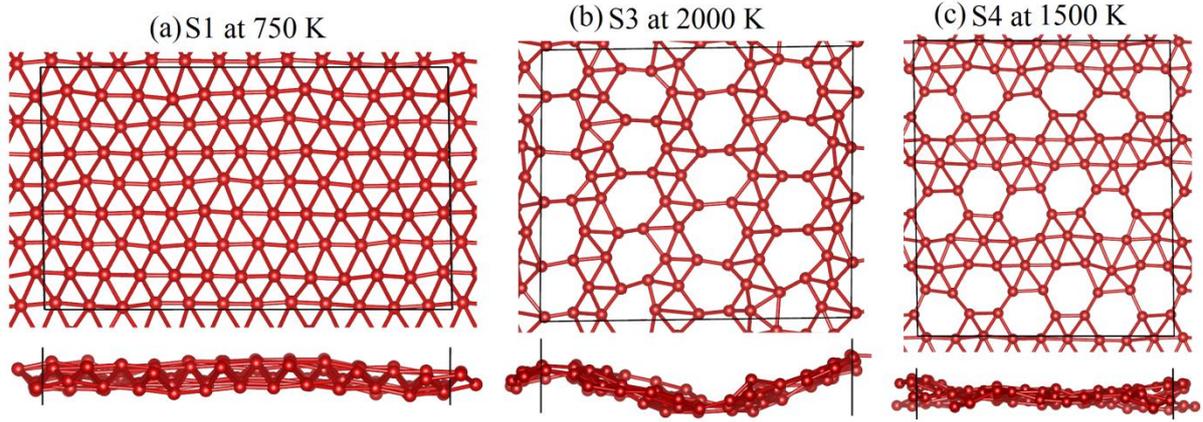

Fig. 8- Thermal stability of borophene films. Snapshots of S1, S3 and S4 borophene sheets at different temperatures obtained using AIMD calculations for 5 ps. VESTA[46] package was utilized to illustrate these structures.

Thermal stability is a desirable quality in nanomaterials of practical use. In this work we also studied the thermal stabilities of the five borophene structures at high temperature using AIMD simulations. 5 picoseconds of simulations were performed for all the structures. The S1 structure remained intact at the end of the simulation at T=750 K. It was partly disintegrated at T=1000 K and T=1500 K and completely disintegrated at T=2000 K. The S2 structure was completely disintegrated at T=3000 K but only partially so at T=2000 K. S3 and S5 Structures were intact at T=2000 K but disintegrated at T=3000 K. The S4 structure was intact at T=1500K but disintegrated at T=2000K. In Fig. 8 sample of S1, S2 and S4 borophene sheets at different temperatures obtained using AIMD calculations for 5 ps are illustrated. Despite of considerable deformation of the structures due to the thermal fluctuations, the chemical bonds were intact which confirms the stability of structures. Putting together, it can be concluded that S1 structure is the least thermally stable among all the structures and none of the structures can withstand a high temperature like T=3000K.

## 4. Summary

We performed extensive first-principles density functional theory calculations to provide a general viewpoint concerning the mechanical properties of five different borophene sheets. To this aim, we applied uniaxial tension condition to study the mechanical properties of borophene sheets. For all the considered borophene structures, we analyzed the effects of loading direction and point vacancy on the mechanical response. Our first-principles modelling revealed that borophene films depending on the boron atoms arrangements and the loading direction can yield remarkable elastic modulus in a range of 163-382 GPa.nm and high ultimate tensile strength from 14 GPa.nm to around 22.8 GPa.nm at a corresponding strain from 0.1 to 0.21. While the elastic modulus and ultimate tensile points of borophene sheets were found to be anisotropic, their Poisson's ratios were predicted to be almost independent of the loading direction. Based on our modelling results the Poisson's ratio of borophene films can vary from -0.01 up to around 0.26. Our simulations results for all relaxed and uniaxially strained systems up to the ultimate tensile strength point, suggest that borophene sheets present metallic behaviour as indicated by the lack of band gap opening in the electronic DOS. In addition, in order to understand the intensity of defects effect on the mechanical properties of borophene sheets, we studied the elastic modulus of borophene membranes with different point defects concentration. We found that the elastic modulus of considered borophene films along the armchair direction decreases almost linearly by increasing the defects concentration. Interestingly, our DFT calculations revealed that the elastic modulus of a particular borophene structure along the zigzag direction slightly increases with increasing the defects concentration. Ab initio molecular dynamics simulations suggests that borophene films depending on their atomic arrangements can withstand temperatures from 750 K to 2000 K. The information provided by the present investigation can be useful to validate the parameterization of force fields for simulation of borophene films at a larger scale using the classical molecular dynamics method.

## Acknowledgment

BM, OR and TR greatly acknowledge the financial support by European Research Council for COMBAT project (Grant number 615132).